\documentclass[12pt, a4paper]{article}      
\usepackage[dvips]{graphicx}
\usepackage{wrapfig}
\usepackage{amssymb}
\usepackage{amsmath}
\usepackage[a4paper,hmargin={3cm,.9in},height=10in]{geometry}
\usepackage{mathrsfs}
\usepackage{eurosym}
\usepackage{dsfont}
\begin{document}

\title{Quantum repeated games revisited
}


\author{Piotr Fr\c{a}ckiewicz}
\newtheorem{lemma}{Lemma}
\newtheorem{definition}{Definition}[section]
\newtheorem{theorem}{Theorem}
\newtheorem{proposition}{Proposition}[section]
\newtheorem{example}[proposition]{Example}
\newtheorem{algorithm}[proposition]{Algoritm}
\newenvironment{xx}{\noindent\textbf{Proof.}}
{\nolinebreak[4]\hfill$\square$\\\par}

\author{\textsc{Piotr Fr\c{a}ckiewicz} \\Institute of Mathematics of the Polish Academy of Sciences\\
00-956 Warsaw, Poland}

\maketitle

\begin{abstract}
We present a scheme for playing quantum repeated $2\times2$ games
based on the Marinatto and Weber's approach \cite{marinatto} to
quantum games. As a potential application, we study twice repeated
Prisoner's Dilemma game. We show that results not available in
classical game can be obtained when the game is played in the
quantum way. Before we present our idea, we comment on the
previous scheme of playing quantum repeated games proposed in
\cite{iqbalrepeat}. We point out the drawbacks that make results
in \cite{iqbalrepeat} unacceptable. \\

\end{abstract}
\section{Introduction}
The Marinatto-Weber (MW) idea of quantum $2\times2$ games
introduced in \cite{marinatto} has found application in many
branches of game theory. The MW approach to evolutionary games
\cite{aspect} and Stackelberg equilibrium \cite{iqbalbackwards}
are merely two of many applications. In the papers
\cite{fracornormalrepresentation} and \cite{imperfect} we have
shown that the MW idea is applicable as well to finite games in
extensive form. Consequently, this scheme of playing quantum games
can be applied to many other game-theoretical problems. In this
paper we deal with the problem of quantization of twice repeated
$2\times2$ games. Since a finitely repeated game is just
a~particular case of a finite extensive game, we apply the method
based on \cite{fracornormalrepresentation} and \cite{imperfect} to
play the repeated game in the quantum way. The idea of quantum
repreated games was first introduced in \cite{iqbalrepeat}, where
the Authors adapt the MW scheme for the twice repeated Prisoner's
Dilemma. Then, they investigate if results that are unavailable
when the game is played classically can occur in the quantum area.
The point of the paper \cite{iqbalrepeat} is to provide sufficient
conditions for players' cooperation in the Prisoner's Dilemma
game. We examine the idea of \cite{iqbalrepeat} before we define
our scheme. Firstly, we study the problem of cooperation
considered by the Authors of \cite{{iqbalrepeat}} and we prove
that player's cooperation in the game defined by the protocol
proposed in \cite{iqbalrepeat} is not possible. Secondly, we check
whether that scheme is actually in accordance with the concept of
repeated game. As we will show, the discussed scheme does not
include the classical twice repeated Prisoner's Dilemma, hence it
cannot be the quantum realization of this game in the spirit of
the MW approach. To support our arguments we propose the new
protocol for a twice repeated $2 \times 2$ game and prove that our
idea generalizes the classical twice repeated game. Our paper also
contains the proof of the advantage of the quantum scheme over the
classical one: We prove that both players can benefit from playing
game via our protocol. Moreover, we show that contrary to the
situation encountered in the classical game, the cooperation of
the players is possible for some sort of Prisoner's Dilemma games
played repeatedly when our quantum approach is used. \vspace{12pt}

\noindent Studying our paper requires little background in game
theory. All notions like extensive game, information set,
strategy, equilibrium, subgame perfect equilibrium etc. used in
the paper are explained in an accessible way, for example, in
\cite{osborne} and \cite{peters}. The adequate preliminaries can
also be found in the paper \cite{fracornormalrepresentation},
where quantum games in an extensive form are examined.
\section{Twice repeated games and the Prisoner's
Dilemma}\label{sectiondwa} 
\begin{figure}[t]
\centering
\includegraphics[angle=0, scale=0.85]{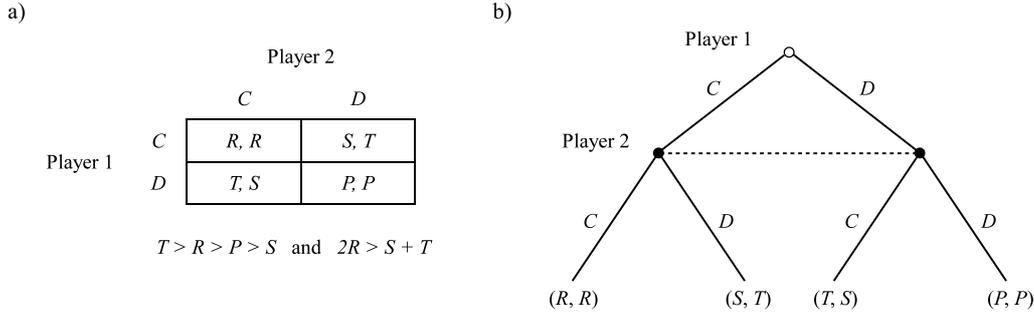}
\caption{The strategic a) and extensive b) form of the Prisoner's
Dilemma.} \label{figure1}
\end{figure}

The Prisoner's Dilemma (PD) is one of
the most fundamental problems in game theory (the general form of
the PD according to \cite{rapoport} is given in
Fig.~\ref{figure1}(a).
It demonstrates why the rationality of players can lead them to an
inefficient outcome. Although the payoff vector $(R,R)$ is better
to both players than $(P,P)$, they cannot obtain this outcome
since each player's strategy $C$ (cooperation) is strictly
dominated by $D$ (defection). As a result, the players end up with
payoff $P$ corresponding to the unique Nash equilibrium $(D,D)$.

\begin{figure}[t]
\centering
\includegraphics[angle=0, scale=0.85]{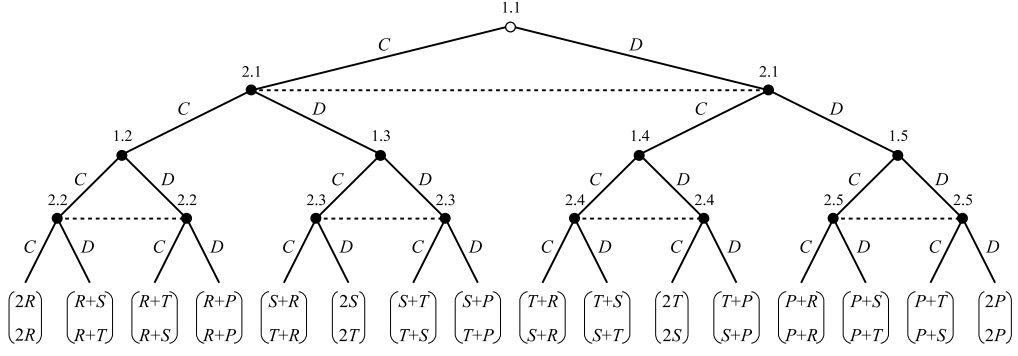}
\caption{The extensive form for a twice repeated Prisoner
Dilemma.} \label{figure2}
\end{figure}

A similar scenario occurs in a case of finitely repeated PD. The
concept of a finitely repeated game assumes playing a static game
(a stage of the repeated game) for a fixed number of times.
Additionally, the players are informed about results of
consecutive stages. In the twice repeated case it means that each
player's strategy specify an action at the first stage and four
actions at the second stage where a particular action is chosen
depending on what of the four outcomes of the first stage has
occurred. It is clearly visible when we write twice repeated game
in the extensive form (see Fig.~\ref{figure2}).
The first stage of the twice repeated PD in the extensive form is
simply the game depicted in Fig.~\ref{figure1}(b) where the
players specify an action $C$ or $D$ at the information set 1.1
and 2.1, respectively (the information sets of player 2 are
distinguished by dotted line connecting the nodes to show lack of
knowledge of the second player about the previous move of the
first player). When the players choose their actions, the result
of the first stage is announced. Since they have knowledge about
the results of the first stage, they can choose different actions
at the second stage depending on the previous result, hence the
next four game trees from Fig.~\ref{figure1} are required to
describe the repeated game. The game tree exhibits ten information
sets (five for each player labelled $1.\cdot$ and $2.\cdot$,
respectively) at which each of the players has two moves. Thus,
each of them has $2^5 = 32$ strategies as they specify $C$ or $D$
at their own five information sets.

To find the Nash equilibrium in finitely repeated game it is
convenient to use the property that the equilibrium profile always
implies the Nash equilibrium at the last stage of the game.
Therefore, to find the Nash equilibrium in the twice repeated PD
it is sufficient to consider strategy profiles that generate the
profile $(D,D)$ at the second stage. Then it follows that $D$ is
the best response for players at the first stage as well. By
induction it can be shown that playing the action $D$ at each
stage of finitely repeated PD constitutes the unique Nash
equilibrium. It is worth noting that if a single stage of repeated
game has more than one equilibrium, different Nash equilibria may
be played at the last stage depending on results of previous
stages. For example, let us consider the Battle of the Sexes (BoS)
game given by the following bimatrix:
\begin{equation}
\begin{array}{l} ~~~~~~~~~~~~~ \quad O  \qquad~~  F \\
\Gamma: \begin{array}{c} O \\ F \end{array}
\left[\begin{array}{cc}
(\alpha,\beta) & (\gamma,\gamma) \\
(\gamma,\gamma) & (\beta,\alpha) \\
\end{array}\right], \quad $where$ \quad \alpha
> \beta > \gamma.
\end{array}
\end{equation}
It has two pure Nash equilibria, namely, $(O,O)$ and $(F,F)$. Let
us examine now the twice repeated BoS. Obviously, its game tree is
the same as one in Fig.~\ref{figure2}. Let us assign appropriate
sum of two stage payoff outcomes to each possible profile (like it
has been done in the case of the twice repeated PD). Then we find
many different Nash equilibria. One of these is to play the Nash
equilibrium $(O,O)$ at the first stage, keep playing $(O,O)$ at
the second stage if the outcome of the first one is $(O,O)$ or
$(O,F)$, otherwise to play stage-game Nash equilibrium $(F,F)$.
\section{Comment on `Quantum repeated games' by Iqbal and Toor \cite{iqbalrepeat}}
Let us remind the MW approach to playing the PD repeatedly
introduced in \cite{iqbalrepeat}. According to this concept the
two-stage PD is placed in $\mathscr{H} =
\left(\mathds{C}^2\right)^{\otimes4}$ complex Hilbert space with
the computational basis. The game starts with preparing 4-qubit
pure state represented by a unit vector in $\mathscr{H}$. The
general form of this state is described as follows:
\begin{align} \begin{split}
&|\psi_{\mathrm{in}}\rangle = \sum_{x_{1}, x_{2}, x_{3}, x_{4} =
0,1} \lambda_{x_{1}, x_{2}, x_{3},
x_{4}}|x_{1},x_{2},x_{3},x_{4}\rangle,\\
&\mbox{where} \quad \lambda_{x_{1}, x_{2}, x_{3}, x_{4}} \in
\mathds{C} \quad \mbox{and} \quad \sum_{x_{1}, x_{2}, x_{3}, x_{4}
= 0,1} |\lambda_{x_{1}, x_{2}, x_{3}, x_{4}}|^2 = 1.
\label{drinitialstate} \end{split}
\end{align}  Players' moves are identified with the identity operator $\sigma_{0}$
and the bit flip Pauli operator $\sigma_{1}$. Player 1 is allowed
to act on the first and third qubit, and player 2 acts on the
second and fourth one.
 In the first stage of the game the two first qubits
are manipulated. Let $\rho_{\mathrm{in}}$ be the density operator
for the initial state (\ref{drinitialstate}). Then the state
$\rho$ after the players' actions takes the form
\begin{align} \label{samoro} \begin{split}
&\rho = \sum_{\kappa_{1}, \kappa_{2} =
0,1}p_{\kappa_{1}}q_{\kappa_{2}}(\sigma^1_{\kappa_{1}} \otimes
\sigma^2_{\kappa_{2}}) \rho_{\mathrm{in}}
(\sigma^1_{\kappa_{1}} \otimes \sigma^2_{\kappa_{2}}),\\
&\mbox{and} \;\; \sum_{\kappa_{1} = 0,1}p_{\kappa_{1}} =
\sum_{\kappa_{2} = 0,1}q_{\kappa_{2}} =1, \end{split}
\end{align}
where $p_{\kappa_{1}}$ ($q_{\kappa_{2}}$) is the probability of
applying $\sigma^1_{\kappa_{1}}$ ($\sigma^2_{\kappa_{2}})$ to the
first (second) qubit. Next, the other two qubits  are manipulated
which, according to Iqbal and Toor, corresponds to the second
stage of the classical game. The operation $\sigma^3_{\kappa_{3}}$
on the third qubit with probability $p_{\kappa_{3}}$ and operation
$\sigma^4_{\kappa_{4}}$ on the fourth qubit with probability
$q_{\kappa_{4}}$ change the state $\rho$ to
\begin{align}\label{finalro} \begin{split}
&\rho_{\mathrm{fin}} = \sum_{\kappa_{3}, \kappa_{4} =
0,1}p_{\kappa_{3}}q_{\kappa_{4}}(\sigma^3_{\kappa_{3}} \otimes
\sigma^4_{\kappa_{4}}) \rho(\sigma^3_{\kappa_{3}} \otimes
\sigma^4_{\kappa_{4}}),
\\ &\mbox{and} \quad \;\; \sum_{\kappa_{3} = 0,1}p_{\kappa_{3}} =
\sum_{\kappa_{4} = 0,1}q_{\kappa_{4}} =1.
\end{split}
\end{align}
The next step is to measure the final state $\rho_{\mathrm{fin}}$
in order to determine final payoffs. The measurement is defined by
the four payoffs operators $X_{i.j}$, $i,j=1,2$ associated with
particular: player $i$ and stage $j$. That is
\begin{align}\label{drpayoffoperator} \begin{split}
X_{1.1} = (R|00\rangle \langle00| + S|01\rangle \langle01| +
T|10\rangle \langle 10| + P|11\rangle \langle 11|) \otimes
\mathds{1}^{\otimes 2};\\
X_{1.2} =\mathds{1}^{\otimes 2} \otimes (R|00\rangle \langle00| +
S|01\rangle \langle01| + T|10\rangle \langle 10| + P|11\rangle
\langle 11|);\\ X_{2.1} = (R|00\rangle \langle00| + T|01\rangle
\langle01| + S|10\rangle \langle 10| + P|11\rangle \langle 11|)
\otimes \mathds{1}^{\otimes 2};\\
X_{2.2} = \mathds{1}^{\otimes 2} \otimes (R|00\rangle \langle00| +
T|01\rangle \langle01| + S|10\rangle \langle 10| + P|11\rangle
\langle 11|).  \end{split}
\end{align}
Then the expected payoff $E_{i.j}$ for player $i$ at stage $j$
when player 1 chooses strategy $(\sigma^1_{\kappa_{1}},
\sigma^3_{\kappa_{3}})$ and player 2 chooses
$(\sigma^2_{\kappa_{2}}, \sigma^4_{\kappa_{4}})$ is obtained by
the following formula:
\begin{align}
E_{i.j}\left((\sigma^1_{\kappa_{1}}, \sigma^3_{\kappa_{3}}),
(\sigma^2_{\kappa_{2}}, \sigma^4_{\kappa_{4}})\right) =
\mathrm{tr}(X_{i.j} \rho_{\mathrm{fin}}). \label{drpayoff}
\end{align}

\noindent The authors took up the issue of cooperation in
two-stage PD. Given the initial state
\begin{align} \label{initialstateiqbal}
|\psi_{\mathrm{in}}\rangle = \lambda_{0000}|0000\rangle +
\lambda_{0011}|0011\rangle + \lambda_{1100}|1100\rangle +
\lambda_{1111}|1111\rangle
\end{align}
and fixed payoffs
\begin{align}\label{pdpayoff}
T = 5, \quad R = 3, \quad P = 1, \quad S = 0,
\end{align}
they identify $\sigma_{0}$ and $\sigma_{1}$ as actions of
cooperation and defection, respectively, and claim that conditions
\begin{align}
|\lambda_{0000}|^2 + |\lambda_{0011}|^2 \leq \frac{1}{3}, \quad
|\lambda_{0011}|^2 + |\lambda_{1111}|^2 \leq \frac{1}{3}
\label{wrongcondition}
\end{align}
are sufficient to choose $\sigma_{0}$ by both players (thereby
cooperating) at the first stage given that the players have chosen
$\sigma_{1}$ at the second one. We raise below two objections
concerning the results of the paper \cite{iqbalrepeat}.
\subsection{The incompatibility of the protocol (\ref{drinitialstate})-(\ref{drpayoff}) and theory of repeated games}
The main fault of the protocol
(\ref{drinitialstate})-(\ref{drpayoff}) is that the twice repeated
game cannot be described in this way. In fact, it quantizes the
game PD played twice when the players are not informed about a
result of the first stage. It is noticeable, for example, when we
re-examine the way of finding the optimal solution provided in
\cite{iqbalrepeat}. The authors analyze the game backwards, first
by focusing on the Nash equilibria at the second stage. They set
condition for the profile $(\sigma^3_{1}, \sigma^4_{1})$ to be the
Nash equilibrium at the second stage. Next, given that
$(\sigma^3_{1}, \sigma^4_{1})$ is fixed, they determine the set of
amplitudes for which the profile $((\sigma^1_{0}, \sigma^2_{0}),
(\sigma^3_{1}, \sigma^4_{1}))$ is the Nash equilibrium of the game
implied by (\ref{drinitialstate})-(\ref{drpayoff}). This method to
find the Nash equilibria is not correct since it doesn't include
the possibility that players make their actions depending on a
result of the first stage. Although the problem seems to be
insignificant where a stage of a repeated game has unique Nash
equilibrium, it becomes visible in remaining cases. Let us
consider the initial state (\ref{initialstateiqbal}) satisfying
the requirement
\begin{align}
2(|\lambda_{0000}|^2 + |\lambda_{1100}|^2) = |\lambda_{0011}|^2 +
|\lambda_{1111}|^2
\end{align}
and let us take (\ref{pdpayoff}) to be PD's payoffs. Then, the
expected payoffs for the players at the second stage of the game
defined by the scheme (\ref{drinitialstate})-(\ref{drpayoff}) are
as follows:
\begin{align}\label{2stage}\begin{split}
&E_{i.2}\bigl((\cdot, \sigma^3_{0}), (\cdot, \sigma^4_{0})\bigr) =
E_{1.2}\bigl((\cdot, \sigma^3_{1}), (\cdot, \sigma^4_{0})\bigr) =
E_{2.2}\bigl((\cdot, \sigma^3_{0}), (\cdot, \sigma^4_{1})\bigr) =
5|\lambda|;\\
&E_{1.2}\bigl((\cdot, \sigma^3_{0}), (\cdot, \sigma^4_{1})\bigr) =
E_{2.2}\bigl((\cdot, \sigma^3_{1}), (\cdot, \sigma^4_{0})\bigr) =
10|\lambda|;\\
&E_{i.2}\bigl((\cdot, \sigma^3_{1}), (\cdot, \sigma^4_{1})\bigr) =
7|\lambda|,
\end{split}
\end{align}
where $|\lambda| = |\lambda_{0000}|^2 + |\lambda_{1100}|^2$ and $i
= 1,2$. Results of (\ref{2stage}) imply continuum of Nash
equilibria in the second stage (it is easy to note, for example,
when we draw a $2\times2$ bimatrix with entries defined by
(\ref{2stage})), among them $\left((\cdot, \sigma^3_{0}),(\cdot,
\sigma^4_{1})\right)$ and $\left((\cdot, \sigma^3_{1}),(\cdot,
\sigma^4_{0})\right)$. Bearing in mind the remark in Section
\ref{sectiondwa} about possible profiles in the BoS game, the
correct protocol for quantum repeated games should be able to
assign a payoff outcome (by the measurement
(\ref{drpayoffoperator})) to a strategy profile, where different
Nash equilibria are played at the second stage depending on
actions chosen at the first one. However, an example of a profile
where the players play $\left((\cdot, \sigma^3_{0}),(\cdot,
\sigma^4_{1})\right)$ at the second stage if a result of the first
stage is $\left((\sigma^1_{0},\cdot),
(\sigma^2_{0},\cdot)\right)$, and they play $\left((\cdot,
\sigma^3_{1}),(\cdot, \sigma^4_{0})\right)$ in other cases cannot
be measured by the scheme (\ref{drinitialstate})-(\ref{drpayoff}).
Since there is two qubit register allotted to the second stage, it
allows to write only one pair of actions $(\sigma^3_{\kappa_{3}},
\sigma^4_{\kappa_{4}})$ before the measurement is made.

An argument against the scheme in \cite{iqbalrepeat} can be
expressed in another way. Namely, all results included in
\cite{iqbalrepeat} can be obtained by considering simplified
protocol (\ref{drinitialstate})-(\ref{drpayoff}) where the
sequential procedure (\ref{samoro}) and (\ref{finalro}) for
determining the final state $\rho_{\mathrm{fin}}$ is simply
replaced with
\begin{align}
\rho_{\mathrm{fin}} = \bigotimes^4_{j=1}\sigma^j_{\kappa_{j}}
\rho_{\mathrm{in}}\bigotimes^4_{j=1}\sigma^j_{\kappa_{j}},
\label{drfinalstate2}
\end{align}
In this case, the first and the second player simultaneously pick
$(\sigma^1_{\kappa_{1}}, \sigma^3_{\kappa_{3}})$ and
$(\sigma^2_{\kappa_{2}}, \sigma^4_{\kappa_{4}})$, respectively,
having essentially only four strategies each. However, as we
mentioned in the previous section, each player has 32 strategies
in the classical twice repeated game. As a result, the protocol
(\ref{drinitialstate})-(\ref{drpayoff}) cannot coincide with the
classical case if $|\psi_{\mathrm{in}}\rangle = |0000\rangle$.
Despite the fact that the Authors assume that a player knows her
opponent's action taken previously, the scheme
(\ref{drinitialstate})-(\ref{drpayoff}) does not take it into
consideration. In consequence, a game being quantized by
(\ref{drinitialstate})-(\ref{drpayoff}) differs from the game in
Fig.~\ref{figure2} in that the nodes 1.2, 1.3, 1.4 and 1.5 (2.2,
2.3, 2.4 and 2.5) lie at the same information set (are connected
with dotted line).
\subsection{The misconception about the cooperative strategy in the PD played via the MW approach}
The another fault, we are going to discuss, is based on
misinterpreting the operators $\sigma_{0}$ and $\sigma_{1}$ as
cooperation and defection in the protocol given by
(\ref{drinitialstate})-(\ref{drpayoff}). Let us consider the
initial state $|\psi_{\mathrm{in}}\rangle$ where the two first
qubits associated with the first stage are prepared in the state
$|x_{1},x_{2}\rangle$, for $x_{1},x_{2} \in \{0,1\}$. Then the
first stage of the game given by
(\ref{drinitialstate})-(\ref{drpayoff}) is isomorphic to the
classical PD game. When the initial state is $|00\rangle$ then
$\sigma_{0}$ corresponds to the action~$C$ and $\sigma_{1}$
corresponds to $D$. However, when the initial state is
$|11\rangle$, the action `cooperate' are identified with
$\sigma_{1}$ and the action `defect' with $\sigma_{0}$ since by
putting $\rho_{\mathrm{fin}} = (\sigma^1_{\kappa_{1}} \otimes
\sigma^2_{\kappa_{2}}) |11 \rangle \langle 11|
(\sigma^1_{\kappa_{1}} \otimes \sigma^2_{\kappa_{2}})$ into the
formula (\ref{drpayoff}) we have
\begin{align}
(\mathrm{tr}(X_{1.1} \rho_{\mathrm{fin}}), \mathrm{tr}(X_{2.1}
\rho_{\mathrm{fin}})) =
\left\{\begin{array}{lll} (R,R), & \mbox{if} & (\kappa_1, \kappa_2) = (1,1);\\
(S,T), & \mbox{if} & (\kappa_1, \kappa_2) = (1,0);\\
(T,S), & \mbox{if} & (\kappa_1, \kappa_2) = (0,1);\\
(P,P), & \mbox{if} & (\kappa_1, \kappa_2) = (0,0).
\end{array}\right. \label{prostyprzyklad}
\end{align}
That is, the outcome of the game does not depend intrinsically on
the operators but depends on the initial state and on what the
final state $\rho_{\mathrm{fin}}$ can be obtained through the
available operators. Thus identification of operators with actions
taken in classical game without taking into consideration the form
of the initial state is not correct. The misidentification assumed
in \cite{iqbalrepeat} implies that the condition
(\ref{wrongcondition}) cannot solve the problem formulated in this
paper. It is clearly visible when we take, for example, the
initial state $|\psi_{\mathrm{in}}\rangle = |1100\rangle$. It
satisfies the inequalities (\ref{wrongcondition}) thus,
$\sigma_{0}$ is optimal at the first stage for each player.  In
fact, $\sigma_{0}$ is the action `defect' as it is shown in
(\ref{prostyprzyklad}). Note also that the payoff corresponding to
the profile $(\sigma_{0}, \sigma_{0})$ at the first stage and
$(\sigma_{1}, \sigma_{1})$ at the second one is $2P$ for each
player - total payoff for the defection. Thus, the condition
(\ref{wrongcondition}) does not ensure the cooperation at the
first stage.

Quite the opposite, it turns out that the players never cooperate
when they play the game defined by
(\ref{drinitialstate})-(\ref{drpayoff}). Let us consider any
initial state (\ref{drinitialstate}) in which the first and the
second qubit are prepared in a way that for $(s_{1},
s_{2})=((\sigma^1_{\kappa_{1}}, \sigma^3_{\kappa_{3}}),
(\sigma^2_{\kappa_{2}}, \sigma^4_{\kappa_{4}}))$ we have
\begin{align}
(E_{1.1}(s_{1}, s_{2}), E_{2.1}(s_{1}, s_{2})) =
\left\{\begin{array}{lll} (R',R'), & \mbox{if} & (\kappa_1, \kappa_2) = (0,0);\\
(S',T'), & \mbox{if} & (\kappa_1, \kappa_2) = (0,1);\\
(T',S'), & \mbox{if} & (\kappa_1, \kappa_2) = (1,0);\\
(P',P'), & \mbox{if} & (\kappa_1, \kappa_2) = (1,1).
\end{array}\right. \label{prostyprzyklad2}
\end{align}
where the values $T', R', P', S'$ meet the requirements of the PD
given in Fig.\ref{figure1}(a), so the operators $\sigma_{0}$ and
$\sigma_{1}$ can be regarded as cooperation and defection,
respectively. Next, let us estimate the difference
\begin{align}
E_{1}((\sigma^1_{1}, \sigma^3_{0}), s_{2}) - E_{1}((\sigma^1_{0},
\sigma^3_{0}), s_{2}) \quad \mbox{for any}\quad s_{2} =
(\sigma^2_{\kappa_{2}}, \sigma^4_{\kappa_{4}}),
\label{drdifference}
\end{align}
where $E_{1} = E_{1.1}+E_{1.2}$. Since the same actions are taken
on the third and the fourth qubit, we have $E_{1.2}((\sigma^1_{0},
\sigma^3_{0}), s_{2}) = E_{1.2}((\sigma^1_{1}, \sigma^3_{0}),
s_{2})$, therefore, the value $E_{1}$ depends only on $E_{1.1}$,
Thus, for $s_{2} = (\sigma^2_{\kappa_{2}},
\sigma^4_{\kappa_{4}})$, we obtain from (\ref{prostyprzyklad2})
that
\begin{align}
0<E_{1}((\sigma^1_{1}, \sigma^3_{0}), s_{2}) -
E_{1}((\sigma^1_{0}, \sigma^3_{0}),
s_{2}) =\left\{\begin{array}{lll}  T'-R', & \mbox{if} & \kappa_{2} = 0;\\
P'-S', & \mbox{if} & \kappa_2 = 1.
\end{array}\right.
\end{align}
In similar way we can prove that the strategy $(\sigma^1_{0},
\sigma^3_{1})$ of player 1 is strictly dominated by
$(\sigma^1_{1}, \sigma^3_{1})$. As a result, we conclude that
$\sigma^1_{1}$ is the best response of player 1 at the first
stage. Symmetry of payoffs in PD implies that strategy
$(\sigma^2_{0}, \sigma^4_{0})$ of player 2 is strictly dominated
by $(\sigma^2_{1}, \sigma^4_{0})$, as well as $(\sigma^2_{0},
\sigma^4_{1})$ is strictly dominated by $(\sigma^2_{1},
\sigma^4_{1})$. Thus, there is no Nash equilibrium in which the
players choose $\sigma_{0}$ (cooperation) at the first stage.
\section{The MW approach to twice repeated quantum games}
In this section we propose a scheme of playing a twice repeated
$2\times2$ quantum game that is free from the faults we have
pointed in the previous section. Our construction is based on the
protocol that we proposed in \cite{fracornormalrepresentation}
where general finite extensive quantum games were considered.
Since a repeated game is a special case of an extensive game, we
can adapt this concept. Next, we examine what results can be
obtained from such protocol. In particular, we re-examine the
problem of cooperation studied in \cite{iqbalrepeat}.
\subsection{Construction of a twice repeated $\mathbf{2\times2}$ quantum game via the MW protocol}

\noindent Let us consider a $2 \times 2$ game defined by the
outcomes $O_{\iota_{1}, \iota_{2}}$, $\iota_{1}, \iota_{2} = 0,1$.
The twice repeated $2\times2$ quantum game played according to the
MW approach is as follows:

\noindent Let $\mathscr{H} = \left(\mathds{C}^2\right)^{\otimes
10}$ be a Hilbert space with the computational basis
$\{|x_{1},x_{2},\dots,x_{10}\rangle\}$, $x_{j} = 0,1$. Then the
initial state of the game is a ten-qubit pure state represented by
a~unit vector in the space $\mathscr{H}$:
\begin{align}
|\psi_{\mathrm{in}}\rangle =
\sum^{2^{10}-1}_{x=0}\lambda_{x}|x\rangle,  \quad \mbox{for} \quad
\lambda_{x} \in \mathds{C} \quad \mbox{and} \quad
\sum_{x}|\lambda_{x}|^2 = 1, \label{drpsi}
\end{align}
where the sum is over all possible decimal values of $x =
(x)_{10}=(x_{1}x_{2}\dots x_{10})_{2}$. The players are allowed to
apply operators $\sigma_{0}$ and $\sigma_{1}$. The qubits with odd
indices are manipulated by player 1 and the qubits labelled by
even indices are manipulated by player 2. Such assignment implies
32 possible strategies for each players as they specify five
operations $\sigma^j_{\kappa_{j}}$ (where $j$ and $\kappa_{j}$
indicate qubit number and operation number, respectively) on their
own qubits. We denote a player $i$'s strategy by $\tau_{i} =
(\sigma^i_{\kappa_{i}}, \sigma^{i+2}_{\kappa_{i+2}},
\sigma^{i+4}_{\kappa_{i+4}},
\sigma^{i+6}_{\kappa_{i+6}},\sigma^{i+8}_{\kappa_{i+8}}),$ where
$i=1,2$. The profile $\tau = (\tau_{1}, \tau_{2})$ gives rise to
the final state:
\begin{align}\label{drpureinitialstate}
|\psi_{\mathrm{fin}}\rangle =
\bigotimes^{10}_{j=1}\sigma^j_{\kappa_{j}}|\psi_{\mathrm{in}}\rangle.
\end{align}
If the players each take $\tau^t_{1}$ and $\tau^{t'}_{2}$ with
probability $p_{t}$ and $q_{t'}$, respectively, that corresponds
to the state $|\psi_{\mathrm{fin}}^{t,t'}\rangle$ (defined by
(\ref{drpureinitialstate})) with probability $p_{t}q_{t'}$, then
the final state is the density operator associated with the
ensemble $\{p_{t}q_{t'}, |\psi_{\mathrm{fin}}^{t,t'}\rangle\}$.
That is
\begin{align}
\rho_{\mathrm{fin}} =
\sum_{t,t'}p_{t}q_{t'}|\psi_{\mathrm{fin}}^{t,t'}\rangle \langle
\psi_{\mathrm{fin}}^{t,t'}|.
\end{align}
Till now, a difference between the concept in \cite{iqbalrepeat}
and our protocol lies in the dimension of the space $\mathscr{H}$.
The next difference is clearly visible in a description of
measurement operators. The measurement on $\rho_{\mathrm{fin}}$
that determines an outcome of the game is described by a
collection $\{X_{1}, X_{2.00}, X_{2.01}, X_{2.10}, X_{2.11}\}$,
where its components are defined as follows:
\begin{align}
X_{1} &= \sum_{x_{1}, x_{2} \in \{0,1\}} O_{x_{1},x_{2}}
|x_{1},x_{2}\rangle \langle x_{1},x_{2}| \otimes
\mathds{1}^{\otimes 8}; \label{drduzyiks}\\
\begin{split}
X_{2.00} &= \sum_{x_{3}, x_{4} \in \{0,1\}} O_{x_{3},
x_{4}}|00\rangle \langle 00| \otimes |x_{3}, x_{4} \rangle \langle
x_{3}, x_{4}| \otimes \mathds{1}^{\otimes 6};\\
X_{2.01} &= \sum_{x_{5}, x_{6} \in \{0,1\}} O_{x_{5},
x_{6}}|01\rangle \langle 01| \otimes \mathds{1}^{\otimes 2}
\otimes |x_{5}, x_{6} \rangle \langle x_{5}, x_{6}| \otimes
\mathds{1}^{\otimes
4};\\
X_{2.10} &= \sum_{x_{7}, x_{8} \in \{0,1\}} O_{x_{7},
x_{8}}|10\rangle \langle 10| \otimes \mathds{1}^{\otimes 4}
\otimes |x_{7}, x_{8} \rangle \langle x_{7}, x_{8}| \otimes
\mathds{1}^{\otimes
2};\\
X_{2.11} &= \sum_{x_{9}, x_{10} \in \{0,1\}} O_{x_{9},
x_{10}}|11\rangle \langle 11| \otimes \mathds{1}^{\otimes 6}
\otimes |x_{9}, x_{10} \rangle \langle x_{9}, x_{10}|.
\label{drduzeiksy}\end{split}
\end{align}
Then the expected outcomes: $E_{i.1}$ at the first stage  and
$E_{i.2}$ at the second stage are calculated by using the
following formulae:
\begin{align}
E_{i.1} = \mathrm{tr}(X_{1} \rho_{\mathrm{fin}}), \quad E_{i.2} =
\mathrm{tr}\left(\sum_{\iota_{1}, \iota_{2}}
X_{2.\iota_{1},\iota_{2}} \rho_{\mathrm{fin}}\right).
\label{dreprotocol}
\end{align}
Let us give justification of our construction. Notice that
$2^{10}$ is a minimal dimension of the space $\mathscr{H}$ in
order to play the twice repeated $2\times2$ game. Since a player's
strategy in a twice repeated $2\times2$ game specifies action at
the first stage and at each of four subgames fixed by the outcome
of the first stage, the quantum protocol needs a five-qubit
register to write a player's strategy. The first two qubits are
used to perform operations at the first stage of the repeated
game. Then given the form of $X_{1}$ and strategies of players
restricted to manipulate the first and the second qubit, in fact,
the protocol (\ref{drpsi})-(\ref{dreprotocol}) coincides with the
MW scheme of playing $2 \times 2$ quantum game \cite{marinatto}.
The remaining eight qubits are used to define players' moves at
the second stage. That is, by pairing consecutive qubits from the
third qubit onwards,
 actions at the second stage are defined on appropriate pair of
qubits depending on the outcome at the previous stage. For
example, given the outcome $O_{10}$ has occurred at the first
stage (that is the outcome 10 on the first two qubits has been
measured), the expected outcome $E_{i.2}$ depends only on
operation on $x_{7}$ and $x_{8}$, i.e, $E_{i.2} =
\mathrm{tr}(X_{2.10}\rho_{\mathrm{fin}})$. Then the players play
the second stage in the same way as in the protocol
(\ref{drinitialstate})-(\ref{drpayoff}). However, contrary to the
previous idea, each player specifies her move for each possible
outcome
$O_{x_{1},x_{2}}$.\\

\noindent A game generated by our scheme naturally coincides with
the classical case when appropriate initial state is prepared. We
prove this fact by means of a convenient sequential approach to
(\ref{drpsi})-(\ref{dreprotocol}) provided in the next section.
\subsection{Extensive form of a quantum twice repeated
$\mathbf{2\times2}$ game}

\noindent The protocol (\ref{drpsi})-(\ref{dreprotocol}) allows to
put a game into an extensive form by using a similar method to
what was described in \cite{fracornormalrepresentation}. The
extensive form is obtained through sequential calculating the
final state $\rho_{\mathrm{fin}}$ according to the following
procedure. At first the players manipulate the first pair of
qubits. Then the measurement in the computational basis is made on
these qubits (as a result, an outcome $O_{\iota_{1}, \iota_{2}}$
of the first stage is returned). The measured outcome is sent to
the players. Depending on the measurement outcome $\iota_{1},
\iota_{2}$ that occurs with probability $p(\iota_{1},\iota_{2})$
 the players act on the next pair of qubits: if $\iota_{1},
 \iota_{2}$ is observed then player 1 and player 2 manipulate qubits $2\iota + 3$ and $2\iota + 4$, respectively, where $\iota = (\iota_{1}
\iota_{2})_{2}$ is a decimal representation of a binary number
$\iota_{1}\iota_{2}$. The procedure can be formally described as
follows:
\begin{algorithm}\label{algorithmostatni}
\end{algorithm}

\newcommand*{\dupaI}{1.}
\newcommand*{\temphead}{$(\sigma^1_{\kappa_{1}}\otimes\sigma^2_{\kappa_{2}})|\psi_{\mathrm{in}}\rangle = |\psi\rangle$}
\newcommand*{\pustyI}{}
\newcommand*{\TempHead}{\footnotesize The players perform their operations $\sigma^1_{\kappa_{1}}$ and $\sigma^2_{\kappa_{2}}$ on the initial state $|\psi_{\mathrm{in}}\rangle$.}
\newcommand*{\dupaII}{2.}
\newcommand*{\tempdrugi}{$\to \displaystyle\frac{M_{\iota_{1},\iota_{2}}|\psi\rangle}{\sqrt{\langle \psi|M_{\iota_{1},\iota_{2}}|\psi \rangle}} = |\psi_{\iota_{1}, \iota_{2}}\rangle$}
\newcommand*{\pustyII}{}
\newcommand*{\Tempdrugi}{\footnotesize The first two qubits in the state $\rho$ are measured. The measurement is described by a collection $\{M_{\iota_{1}, \iota_{2}} \colon M_{\iota_{1}, \iota_{2}} = |\iota_{1},\iota_{2} \rangle \langle \iota_{1},\iota_{2}| \otimes \mathds{1}^{\otimes 8},\iota_{1}, \iota_{2} = 0,1\}$.}
\newcommand*{\dupaIII}{3.}
\newcommand*{\temptrzeci}{$\to \{p(\iota_{1},\iota_{2}), (\sigma^{2\iota+3}_{\kappa_{2\iota+3}}~\otimes~\sigma^{2\iota+4}_{\kappa_{2\iota+4}})|\psi_{\iota_{1},\iota_{2}}\rangle\}$ $~~~~~~~p(\iota_{1},\iota_{2}) =
\langle \psi|M_{\iota_{1},\iota_{2}}|\psi \rangle$}
\newcommand*{\pustyIII}{}
\newcommand*{\Temptrzeci}{\footnotesize Given that the outcome $\iota_{1},\iota_{2}$ has been observed,
players 1 and 2 perform operations
$\sigma^{2\iota+3}_{\kappa_{2\iota+3}}$ and
$\sigma^{2\iota+4}_{\kappa_{2\iota+4}}$ on the post-measurement
state.}
\newcommand*{\dupaIV}{}
\newcommand*{\tempczwarty}{}
\newcommand*{\pustyIV}{}
\newcommand*{\Tempczwarty}{}
\setlength{\tabcolsep}{-5mm}
\renewcommand*{\arraystretch}{1.5}
\begin{tabular}{p{2cm}p{6cm}p{3cm}p{7.4cm}}
\dupaI & \temphead & \pustyI & \TempHead\\
\dupaII & \tempdrugi & \pustyII & \Tempdrugi\\
\dupaIII & \temptrzeci & \pustyIII & \Temptrzeci\\
\dupaIV & \tempczwarty & \pustyIV & \Tempczwarty
\end{tabular}

\noindent It turns out that we can prove \vspace{12pt}
\begin{proposition}
The density operator $|\psi_{\mathrm{fin}}\rangle \langle
\psi_{\mathrm{fin}}|$ associated with state
(\ref{drpureinitialstate}) and the density operator for the
ensemble $\{p(\iota_{1},\iota_{2}),
(\sigma^{2\iota+3}_{\kappa_{2\iota+3}} \otimes
\sigma^{2\iota+4}_{\kappa_{2\iota+4}})|\psi_{\iota_{1},\iota_{2}}\rangle\}$
in Algorithm~\ref{algorithmostatni} determine the same outcomes
$E_{i.1}$ and $E_{i.2}$ with regard to the measurement
(\ref{drduzyiks})--(\ref{drduzeiksy}).\end{proposition}
\vspace*{12pt}

\begin{figure}[t]
\centering
\includegraphics[angle=0, scale=0.85]{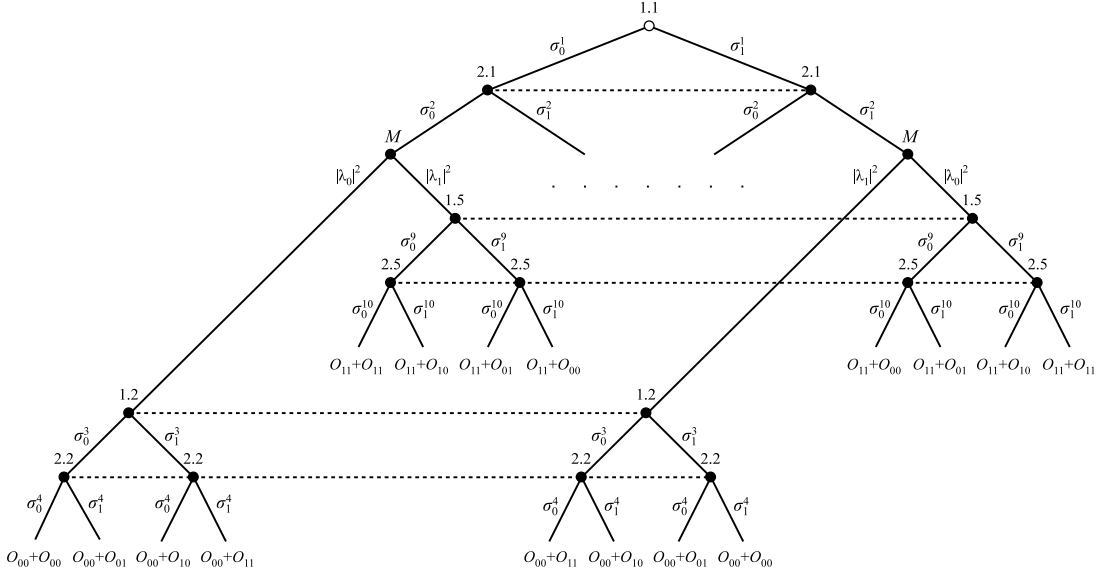}
\caption{The extensive form for a twice repeated Prisoner Dilemma
played through protocol (\ref{drpsi})-(\ref{dreprotocol}) when the
initial state is on the form of (\ref{drsplatany}).}
\label{figure3}
\end{figure}

\begin{xx}
Let us put $\rho = |\psi\rangle \langle \psi|$. Given that
$|\psi_{\iota_{1}, \iota_{2}}\rangle \langle \psi_{\iota_{1},
\iota_{2}}| = M_{\iota_{1},\iota_{2}}\rho
M_{\iota_{1},\iota_{2}}/p(\iota_{1},\iota_{2})$ the state
$\rho'_{\mathrm{fin}}$ can be written as:
\begin{align}
\rho'_{\mathrm{fin}} = \sum_{\iota_{1}, \iota_{2} = 0,1}
\sigma^{2\iota+3}_{\kappa_{2\iota+3}} \otimes
\sigma^{2\iota+4}_{\kappa_{2\iota+4}} M_{\iota_{1},\iota_{2}}\rho
M_{\iota_{1},\iota_{2}}\sigma^{2\iota+3}_{\kappa_{2\iota+3}}
\otimes \sigma^{2\iota+4}_{\kappa_{2\iota+4}}.
\end{align}
Since the first and the second qubits are measured, any operation
$\sigma^j_{\kappa_{j}}$ for which $j \ne 1,2$ does not influence
the measurement. Therefore we have
\begin{align}
\rho'_{\mathrm{fin}} = \sum_{\iota_{1}, \iota_{2} = 0,1}
M_{\iota_{1},\iota_{2}} \sigma^{2\iota+3}_{\kappa_{2\iota+3}}
\otimes \sigma^{2\iota+4}_{\kappa_{2\iota+4}} \rho\,
\sigma^{2\iota+3}_{\kappa_{2\iota+3}} \otimes
\sigma^{2\iota+4}_{\kappa_{2\iota+4}}M_{\iota_{1},\iota_{2}}.
\label{drequality}
\end{align}
Note that $X_{\iota'_{1},\iota'_{2}}M_{\iota_{1},\iota_{2}} =
\delta_{\iota,\iota'}X_{\iota'_{1},\iota'_{2}}$, where
$\delta_{\iota,\iota'}$ is the Kronecker's delta, and $\iota
=(\iota_{1},\iota_{2})_{2}$, and $\iota'
=(\iota'_{1},\iota'_{2})_{2}$, Using the form (\ref{drequality})
of $\rho'_{\mathrm{fin}}$ we have
\begin{align}
\mathrm{tr}\left(\sum_{\iota'}X_{2.\iota'_{1}, \iota'_{2}}
\rho'_{\mathrm{fin}}\right) =
\mathrm{tr}\left(\sum_{\iota}X_{2.\iota_{1},
\iota_{2}}\sigma^{2\iota+3}_{\kappa_{2\iota+3}} \otimes
\sigma^{2\iota+4}_{\kappa_{2\iota+4}} \rho
\sigma^{2\iota+3}_{\kappa_{2\iota+3}} \otimes
\sigma^{2\iota+4}_{\kappa_{2\iota+4}} \right). \label{drequation2}
\end{align}
For each $\iota$ the trace of each term of the sum on the
right-hand side of equation (\ref{drequation2}) depends only on an
operation $\sigma^j_{\kappa_{j}}$ on a qubit $j$, where $j \in
\{1,2,2\iota+3,2\iota+4\}$. Thus, the equation (\ref{drequation2})
holds when also the rest of operations $\sigma^j_{\kappa_{j}}$ are
added:
\begin{align}
\mathrm{tr}\left(\sum_{\iota}X_{2.\iota_{1}, \iota_{2}}
\rho'_{\mathrm{fin}}\right) = \mathrm{tr}\left(\sum_{\iota}
X_{2.\iota_{1},\iota_{2}}\bigotimes^{10}_{j=1}\sigma^j_{\kappa_{j}}\rho_{\mathrm{in}}\bigotimes^{10}_{j=1}\sigma^j_{\kappa_{j}}
\right). \label{drequation3}
\end{align}
As a result, the left-hand side of (\ref{drequation3}) is equal to
the expected outcome $E_{i.2}$ associated with the final state
$|\psi_{\mathrm{fin}}\rangle \langle\psi_{\mathrm{fin}}|$. To
prove that $\rho'_{\mathrm{fin}}$ also determines the expected
outcome $E_{i.1}$ let us see that $X_{1}$ and
$\{M_{\iota_{1},\iota_{2}}\}$ are the same projective measurement
up to the eigenvalues. Hence
\begin{align}
\mathrm{tr}\left(X_{1}\rho'_{\mathrm{fin}}\right) =
\mathrm{tr}\left(X_{1}\sigma^{2\iota+3}_{\kappa_{2\iota+3}}
\otimes \sigma^{2\iota+4}_{\kappa_{2\iota+4}} \rho
\sigma^{2\iota+3}_{\kappa_{2\iota+3}} \otimes
\sigma^{2\iota+4}_{\kappa_{2\iota+4}} \right).\end{align} Since
$\rho = \sigma^1_{\kappa_{1}} \otimes \sigma^2_{\kappa_{2}}
\rho_{\mathrm{in}} \sigma^1_{\kappa_{1}} \otimes
\sigma^2_{\kappa_{2}}$, we obtain
\begin{align}
\mathrm{tr}\left(X_{1} \rho'_{\mathrm{fin}}\right) =
\mathrm{tr}\left(
X_{1}\bigotimes^{10}_{j=1}\sigma^j_{\kappa_{j}}\rho_{\mathrm{in}}\bigotimes^{10}_{j=1}\sigma^j_{\kappa_{j}}
\right). \label{drequation4}
\end{align}
Equations (\ref{drequation3}) and (\ref{drequation4}) show that
the state determined by the sequential procedure and state
(\ref{drpureinitialstate}) set the same outcomes $E_{i.1}$ and
$E_{i.2}$ for $i=1,2$. Using the same way as above and the
linearity of the trace it can be proved that the equivalence is
true if the players pick nondegenerate mixed strategies as well.
\end{xx}
\noindent Having a sequential approach that is in conformity with
protocol (\ref{drpsi})-(\ref{dreprotocol}) we are able to analyze
a quantum repeated game through an extensive form. It can
facilitate the work significantly bearing in mind $32\times32$
bimatrix associated with the normal representation of twice
repeated $2\times2$ game. Let us study the game tree drawn from
the sequential procedure if the initial state (\ref{drpsi}) takes
the form
\begin{align}
|\psi_{\mathrm{in}}\rangle = \lambda_{0}|0\rangle^{\otimes 10} +
\lambda_{1}|1\rangle^{\otimes 10}. \label{drsplatany}
\end{align}
Let us use the sequential procedure step by step. At first the
players manipulate $\sigma^1_{\kappa_{1}}$ and
$\sigma^2_{\kappa_{2}}$. Hence we obtain the following state:
\begin{align}
\sigma^1_{\kappa_{1}} \otimes
\sigma^2_{\kappa_{2}}|\psi_{\mathrm{in}}\rangle =
\lambda_{0}|\kappa_{1},\kappa_{2}\rangle |0\rangle^{\otimes 8} +
\lambda_{1}|\overline{\kappa}_{1},\overline{\kappa}_{2}\rangle
|1\rangle^{\otimes 8}, \label{dr2state}
\end{align}
where $\overline{\kappa}_{j}$ is the negation of $\kappa_{j}$. A
game tree at this phase is just the game tree corresponding to
a~$2\times2$ game (see Fig.~\ref{figure1}(b)), where
$\sigma^j_{\kappa_{j}}$ for $j=1,2$, $\kappa_{j} = 0,1$ are
associated with respective branches of that game tree. After a
sequence of actions $(\sigma^1_{\kappa_{1}},
\sigma^2_{\kappa_{2}})$ the measurement $\{M_{\iota_{1},
\iota_{2}}\}$ is made. Let us focus on the cases when the
measurement outcome 00 or 11 has been observed. The form of
(\ref{dr2state}) tells us that the measurement outcomes 00 and 11
are possible only if the profile at the first stage takes the form
of $(\sigma^1_{\kappa}, \sigma^2_{\kappa})$, where $\kappa = 0,1$.
Then, the probability $p(00)$ ($p(11)$) that the measurement
outcome 00 (11) will occur is equal to $|\lambda_{\kappa}|^2$
($|\lambda_{\overline{\kappa}}|^2$). Thus, the game tree is
extended to include random actions 00 and 11 with associated
probabilities after the both histories $(\sigma^1_{\kappa},
\sigma^2_{\kappa})$. Since further moves of the players depend
only on the measurement, the pair of histories $(\sigma^1_{0},
\sigma^2_{0},00)$, $(\sigma^1_{1}, \sigma^2_{1}, 00)$ and the pair
$(\sigma^1_{0}, \sigma^2_{0},11)$, $(\sigma^1_{1}, \sigma^2_{1},
11)$ constitute two separate information sets. Next, given that 00
(11) has occurred, following the sequential procedure, the players
manipulate third and fourth (ninth and tenth) qubit at the second
stage. Therefore another extensive form of $2\times2$ is added to
each sequence $(\sigma^1_{\kappa}, \sigma^2_{\kappa},\iota\iota)$,
where $\kappa,\iota = 0,1$. In consequence we obtain a game tree
shown in Fig.~\ref{figure3} (a part of the game tree after
histories of $(\sigma^1_{\kappa}, \sigma^2_{\overline{\kappa}})$,
$\kappa = 0,1$ is similar).
Each outcome associated with a terminal sequence are determined by
a pure state from the ensemble given by the sequential procedure.
For example, after sequence $(\sigma^1_{1}, \sigma^2_{1})$ the
post-measurement state takes the form of $|0\rangle^{\otimes
2}|1\rangle^{\otimes 8}$ (up to a global phase factor) with
probability $|\lambda_{1}|^2$, and the players choose sequence
$(\sigma^3_{\kappa_{3}},\sigma^4_{\kappa_{4}})$. Then the total
outcome $E_{i}\mathrel{\mathop:}=E_{i.1} + E_{i.2}$ associated
with a sequence $(\sigma^1_{1}, \sigma^2_{1}, 00,
\sigma^3_{\kappa_{3}},\sigma^4_{\kappa_{4}})$ is calculated
according to formulae (\ref{dreprotocol}):
\begin{align}
E_{i} = \mathrm{tr}\left(\left(X_{1}+ \sum_{\iota_{1},\iota_{2}}
X_{2.\iota_{1},\iota_{2}}\right)(|0\rangle \langle 0|)^{\otimes
2}|\overline{\kappa}_{3},\overline{\kappa}_{4}\rangle \langle
\overline{\kappa}_{3},\overline{\kappa}_{4}|(|1\rangle \langle
1|)^{\otimes 6}\right).
\end{align}
The extensive approach allows us to see directly that our scheme
coincides with the classical twice repeated $2\times2$ game when
$|\psi_{\mathrm{in}}\rangle = |0\rangle^{\otimes 10}$. Without
loss of generality, let the outcomes $O_{\iota_{1},\iota_{2}}$ be
the payoff outcomes corresponding to the PD game. Then putting
$|\lambda_{0}|^2 = 1$ in (\ref{drsplatany}) and assuming
$\sigma^j_{0} \mathrel{\mathop:}= C$, $\sigma^j_{1}
\mathrel{\mathop:}= D$ the game in Fig.~\ref{figure3} depicts
exactly the classical twice repeated PD game (compare Fig.
\ref{figure2} and Fig. \ref{figure3}).
\subsection{Twice repeated PD game played by means of the protocol (\ref{drpsi})-(\ref{dreprotocol})}
Let us study the twice repeated PD game played with the use of our
scheme. Analysis of our protocol with the general form initial
state (\ref{drpsi}) is a laborious task and it deserves a separate
paper to report about. Nevertheless, we can derive many
interesting features with less effort considering the initial
state of the form
\begin{align}
|\psi_{\mathrm{in}}\rangle =
\bigotimes^5_{j=1}|\varphi_{j}\rangle, \;\mbox{where}\;
|\varphi_{j}\rangle\; \mbox{is a state of $2j-1$ and $2j$ qubit}.
\label{drprostyinitialstate}
\end{align}
Let us consider first the problem of optimization of the
equilibrium payoffs,
 given a~space of initial states as a domain.
\vspace*{12pt}
\begin{proposition}
There are infinitely many settings of the initial state
(\ref{drpsi}) for which the twice repeated PD game played with the
use of the protocol (\ref{drpsi})-(\ref{dreprotocol}) has a unique
subgame perfect equilibrium with the equilibrium payoff $(2Q,2Q)$
such that $Q>P$. \label{drproposition}
\end{proposition}
\vspace*{12pt}
\begin{xx}
Let us put the initial state (\ref{drprostyinitialstate}) into the
protocol (\ref{drpsi})-(\ref{dreprotocol}) assuming that
$|\varphi_{j}\rangle = |\varphi\rangle$ for any $j$. Then, the
measurement $\{M_{\iota_{1},\iota_{2}}\}$ on the first pair of
qubits does affect others qubits. Moreover, given that the outcome
$O_{\iota_{1},\iota_{2}}$ has occurred, the expected outcome
$E_{i.2}$ depends only on manipulating on one pair of qubits
$|\varphi\rangle$ due to the form of (\ref{drduzeiksy}).
Therefore, regardless of the first stage outcome
$O_{\iota_{1},\iota_{2}}$, the players are faced with a~$2\times2$
quantum game at the second stage (played via the MW approach).
That is, the players are faced with the problem
\begin{align}
\left(|\varphi\rangle \langle \varphi|, \{\sigma_{0},
\sigma_{1}\}, X'_{i}\right), \label{drpdmw}
\end{align}
where player 1 and 2 apply operators from the set $\{\sigma_{0},
\sigma_{1}\}$ on the first and the second qubit of
$|\varphi\rangle$, respectively. The outcome operator $X'_{i}$
takes the form
\begin{align}
X'_{i} = \sum_{y_{1},y_{2}=0,1} O_{y_{1},y_{2}}|y_{1},y_{2}\rangle
\langle y_{1},y_{2}|,
\end{align}
and the expected outcome is equal to
\begin{align}
E_{i}(\sigma^1_{\kappa_{1}}, \sigma^2_{\kappa_{2}}) =
\mathrm{tr}\left(\sigma^1_{\kappa_{1}} \otimes
\sigma^2_{\kappa_{2}}|\varphi\rangle \langle
\varphi|\sigma^1_{\kappa_{1}} \otimes \sigma^2_{\kappa_{2}}
X'_{i}\right).
\end{align}
Obviously, the first stage game is also described exactly as the
triple (\ref{drpdmw}). Since a~quantum game according to the MW
approach is a game expressed by a bimatrix, it leads us to the
conclusion that protocol (\ref{drpsi})-(\ref{dreprotocol}) with
the initial state $|\varphi \rangle^{\otimes 5}$, in fact, can be
treated as
 a twice repeated bimatrix game generated by (\ref{drpdmw}).

\noindent Let us substitute $O_{y_{1},y_{2}}$ for the payoffs of
the PD game in the game (\ref{drpdmw})
 and examine it towards uniqueness of Nash equilibria. Putting a state
$|\varphi\rangle = \lambda_{0}|00\rangle + \lambda_{1}|11\rangle$,
for which the amplitudes of $|\varphi\rangle$ satisfy the
condition:
\begin{align}
0<|\lambda_{0}|^2 < \frac{\min\{T-R, P-S\}}{T-R+P-S}
\label{drconditionend}
\end{align}
the inequalities
\begin{align}\label{inequalities}
E_{1}\left(\sigma^1_{0},\sigma^2_{\kappa_{2}}\right) >
E_{1}\left(\sigma^1_{1},\sigma^2_{\kappa_{2}} \right) \quad
\mbox{and} \quad
E_{2}\left(\sigma^1_{\kappa_{1}},\sigma^2_{0}\right)
> E_{1}\left(\sigma^1_{\kappa_{1}},\sigma^2_{1} \right)
\end{align}
are true for any $\kappa_{1}, \kappa_{2} = 0,1$. Inequalities
(\ref{inequalities}) imply the unique Nash equilibrium
$(\sigma^1_{0},\sigma^2_{0})$. Moreover, the first inequality of
condition (\ref{drconditionend}) ensures that \begin{align}
E_{1}\left(\sigma^1_{0},\sigma^2_{0}\right) = |\lambda_{0}|^2R +
|\lambda_{1}|^2P
> P.
\end{align} Since the game constructed in the proof can be regarded as a classical
twice repeated game, we are allowed to use all facts of classical
repeated game theory. One of these tells us that a unique
stage-game Nash equilibrium implies, for any finite number of
repetitions, a unique subgame perfect equilibrium in which the
stage-game Nash equilibrium is played in every stage. This
completes the proof.
\end{xx}
Of course, the protocol (\ref{drpsi})-(\ref{dreprotocol}) can be
re-formulated for any finitely repeated $2\times2$ game and then
statement analogical to Proposition \ref{drproposition} can be
articulated. Unfortunately, the number of qubits required in our
protocol grows exponentially with number of stages. For example,
in the case of a game repeated three times, the protocol
(\ref{drpsi})-(\ref{dreprotocol}) needs next 32 qubits to describe
the third stage. In general, the number of $\sum^n_{j=1} 2^{2j-1}$
qubits is required for a $2\times2$ game repeated $n$ times.
\vspace*{12pt}

\noindent We shall re-examine now the problem of cooperation
considered in \cite{iqbalrepeat}. We demonstrated in Section 3
that the cooperation at the first stage is not possible in the
game defined by the Iqbal and Toor scheme. However, we also showed
that this protocol does not take into consideration a player's
move at the second stage as a function of the first stage result.
Therefore, in fact it does not allow to study the cooperation
problem in a~proper way. The following example proves that the
cooperation of players is possible if the twice repeated PD game
is played via our scheme. \vspace{12pt}
\begin{example}
\textup{Let us set the PD game with payoff vectors
\begin{align} O_{00} = (4,4),\; O_{01} = (0,5),\; O_{10} =
(5,0),\; O_{11} = (1,1) \label{drpayoffvectors}\end{align}
inserted in (\ref{drduzyiks}) and (\ref{drduzeiksy}). Let us also
assume that the initial state (\ref{drpsi}) takes the form
\begin{align}
|\psi_{\mathrm{in}}\rangle = |0\rangle^{\otimes
2}\left(\sqrt{0,6}|0\rangle^{\otimes 2} +
\sqrt{0,4}|1\rangle^{\otimes 2} \right)|0\rangle^{\otimes 6}.
\label{drstateexample}
\end{align}
A game specified in this way differs from the classical one only
in the subgame following the outcome $O_{00}$ of the first stage
because then $E_{i.2}$ depends on operations on entangled third
and fourth qubit. Since two first qubits in the state $|00\rangle$
imply the classical PD game at the first stage, we are permitted
to identify the action `cooperate' and the action `defect' with
$\sigma_{0}$ and $\sigma_{1}$, respectively, assuming
$C\mathrel{\mathop:}=\sigma_{0}$ and $D \mathrel{\mathop:}=
\sigma_{1}$. Moreover, the quantum measurement after the first
stage is trivialized in this case and it coincides with the
classical observation in an extensive game. It follows that both
the game defined by (\ref{drpsi})-(\ref{dreprotocol}),
(\ref{drpayoffvectors}), (\ref{drstateexample}) and the classical
game can be represented by the same game tree as well as the same
payoff values except when 00 has been measured on the first pair
of qubits after the first stage. Let us determine now the payoff
outcomes at the second stage given that the post-measurement state
of the first pair of qubits is $|00\rangle$ (in other words, when
player 1's strategy is $\tau_{1} = \left(\sigma^1_{\kappa_{1}},
\sigma^3_{\kappa_{3}}, \sigma^5_{\kappa_{5}},
\sigma^7_{\kappa_{7}}, \sigma^9_{\kappa_{9}}\right)$ and player
2's strategy is $\tau_{2} = \left(\sigma^2_{\kappa_{2}},
\sigma^4_{\kappa_{4}}, \sigma^6_{\kappa_{6}},
\sigma^8_{\kappa_{8}}, \sigma^{10}_{\kappa_{10}}\right)$, which
makes the strategy profile in the form $(\tau_{1}, \tau_{2}) =
\left(\left(\sigma^1_{0}, \cdot, \cdot, \cdot,
\cdot),(\sigma^2_{0},\cdot, \cdot, \cdot, \cdot\right)\right)$).
Given the initial state (\ref{drstateexample}) and the form of
operators (\ref{drduzeiksy}), the payoff outcome $E_{i.2}$ for
each $\kappa_{3}, \kappa_{4} \in \{0,1\}$ and $i=1,2$ is as
follows:
\begin{align}
E_{i.2}\left(\left(\sigma^1_{0}, \sigma^3_{\kappa_{3}}, \cdot,
\cdot, \cdot),(\sigma^2_{0},\sigma^4_{\kappa_{4}}, \cdot, \cdot,
\cdot\right)\right) = 0,6O_{\kappa_{3},\kappa_{4}} +
0,4O_{\overline{\kappa}_{3},\overline{\kappa}_{4}}.
\end{align}
The extensive form of the game with expected payoffs $E_{i.1 } +
 E_{i.2}$ given by (\ref{drpayoffvectors}) is shown in Fig.~\ref{figure4}.
\begin{figure}[t]
\centering
\includegraphics[angle=0, scale=0.85]{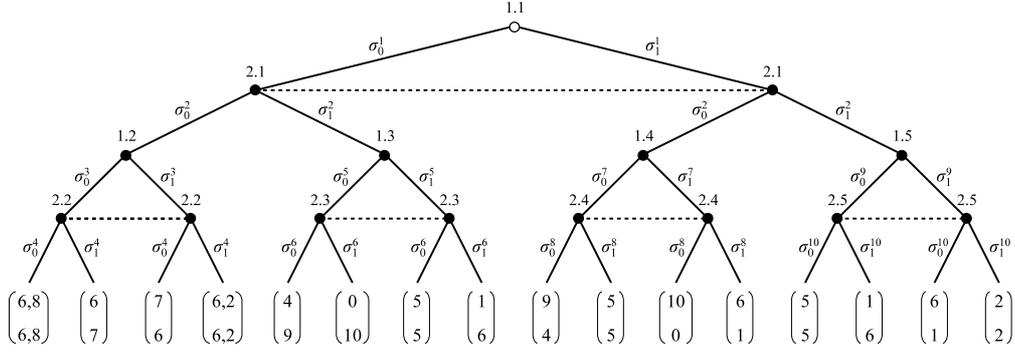}
\caption{The extensive form for a twice repeated Prisoner Dilemma
(\ref{drpayoffvectors}) played through protocol
(\ref{drpsi})-(\ref{dreprotocol}) with update on the initial state
(\ref{drstateexample}).} \label{figure4}
\end{figure}
Let us examine this game for subgame perfect equilibria. Such
profile has to induce the Nash equilibrium in any subgame fixed by
an outcome at first stage. In our case, it is a profile in which
both players take $\sigma_{1}$ on qubits from the third qubit
onward. Consequently, in quest of subgame perfect equilibria, we
take only the following profiles into consideration:
\begin{align}
(\tau_{1}, \tau_{2}) \in \left\{\sigma^1_{\kappa_{1}}\times
\sigma^2_{\kappa_{2}} \times \prod^{10}_{j=3}
\sigma^j_{1}\right\}.
\end{align}
  Then it turns out that the noncooperative
 subgame perfect equilibrium is still preserved. If one of the players picks $\sigma_{1}$ at the first stage,
 the best response of the other one is to pick $\sigma_{1}$ too.
 Therefore, the profile $(\tau'_{1}, \tau'_{2}) =
 \prod^{10}_{j=1}\sigma^j_{1}$ constitutes a subgame perfect
 equilibrium. However, contrary to the classical twice repeated PD, there is another
subgame perfect equilibrium $(\tau''_{1}, \tau''_{2})$ in which
 each player chooses $\sigma_{0}$ (cooperates) at the first stage i.e., $\tau''_{1} = \left(\sigma^1_{0}, \sigma^3_{1}, \sigma^5_{1}, \sigma^7_{1},
 \sigma^9_{1}\right)$ and $\tau''_{2} = \left(\sigma^2_{0}, \sigma^4_{1}, \sigma^6_{1}, \sigma^8_{1},
 \sigma^{10}_{1}\right)$. Moreover, only the latter equilibrium is reasonable since it yields the payoff 6,2 instead of
 2 for each player.} \label{drexample}
\end{example}
\vspace{12pt} Example \ref{drexample} shows that the cooperation
of players is possible when the twice repeated PD game is played
according to our scheme. Unfortunately, the example does not solve
this problem for any PD game. The condition $2R
> T+S$ imposed on the payoffs allows to select an arbitrary large finite number $T$ (if a sufficiently small number $S$ is selected). We suppose that an appropriately large $T$
may convince the players to defect even if the game is played in
quantum domain.
\section{Conclusion}
Our paper proves that repeated games can be quantized. That is, we
have shown that appropriately modified the MW scheme for
$2\times2$ quantum games can indeed generalize a twice repeated
game. In addition, such quantized game can be further analyzed by
strategic as well as extensive form games. Our results also
indicate (with the use of the twice repeated Prisoner's Dilemma)
that playing repeated games in the quantum domain can give
superior results in comparison with the classical ones. At the
same time we have answered why the previous approach
\cite{{iqbalrepeat}} cannot be treated as a correct protocol for
quantum repeated games. The main objection is that the protocol
\cite{{iqbalrepeat}} is unable to consider a full set of
strategies available to players. In contrary to the Iqbal and
Toor's scheme, the protocol defined in this paper is free from the
mentioned fault.

\section*{Acknowledgements} \noindent The author is very grateful to
his supervisor Prof. J. Pykacz from the Institute of Mathematics,
University of Gda\'nsk, Poland for his great help in putting this
paper into its final form.

\end{document}